\documentstyle[11pt,newpasp,twoside,psfig]{article}
\def\edcomment#1{\iffalse\marginpar{\raggedright\sl#1\/}\else\relax\fi}
\reversemarginpar

\begin{document}
\title{Inner density waves: \\ the need for two-dimensional spectroscopy}
 \author{Eric Emsellem}
\affil{Centre de Recherche Astronomique de Lyon, 9 av. Charles Andr\'e, 
	69561 Saint Genis Laval, France}

\begin{abstract}
In this paper, I emphasize the need for integral field spectroscopy
in the study of density waves (bars, spirals, $m=1$ modes) in the centre
of galaxies. I illustrate this by focusing on the complex and time
dependent structures revealed by such techniques.
\end{abstract}

\section{Introduction}

High resolution photometry shows the presence of
non-axisymmetric structures in the inner parts of galaxies:
spirals, bars and lopsided luminosity distributions.
These are assumed to be small versions of the large-scale density
waves observed in disk galaxies. although the involved formation and evolution
processes may be significantly different (see e.g. Emsellem 2002
and references therein). We still do not know the respective
roles of these modes in the evolution of the central kiloparsec, although
possible links with the nuclear activity have been searched for.
More importantly, the dynamical status of these photometric structures
has not been properly assessed: we still don't really know if 
these are true kinematic density waves, or transient density perturbations.
Hence the need for detailed spectroscopic information.

I have discussed some features of inner density waves in a recent paper 
(Emsellem 2002). I wish here to further illustrate why classical long-slit spectroscopy
is not well adapted to unveil the full complexity of the central regions 
of galaxies. These non axisymmetric structures are also strongly time dependent.
This adds an extra dimension to the solution space to probe. 
We then (and always) need all the information we can gather to properly constrain
the morphology and dynamics of these systems.

\section{Where should we put the slit?}

In this Section, I give two prototypical examples where the use
of long-slit spectrography can lead to a restricted and biased
interpretation of the observed structures.

\subsection{A biased view}

One of the best studied nearby nucleus belongs to the spiral galaxy
M~31. With a distance of about 0.7~Mpc ($\sim 3.4$~pc/arcsec), 
it is the closest disk galaxy for which we have a rather 
privileged view of its central structures. HST photometry (Lauer et al. 1993)
has revealed a double nucleus, the brightest peak (P1) being offset by about 
$\sim1.8$~pc from the centre of the outer isophotes (P2). 
Ground-based stellar kinematics (Bacon et al. 1994) 
showed that the velocity field, at a spatial resolution
of 0\farcs9, was nearly symmetric with respect to P2.
The velocity dispersion however peaked on the side opposite to
P1, with respect to P2. Subsequent ground-based long-slit data 
(Kormendy \& Bender 1999, 0\farcs64 FWHM) confirmed the offset of the dispersion peak from
the assumed velocity centre. Statler et al. (1999) then presented 
high resolution FOC long-slit data taken along the P1-P2 axis:
although the signal-to-noise of these data was rather low, they
were able to uncover a strong asymmetry in the stellar velocity
profile, and a high central dispersion value. 
\begin{figure}[h]
\centerline{\psfig{figure=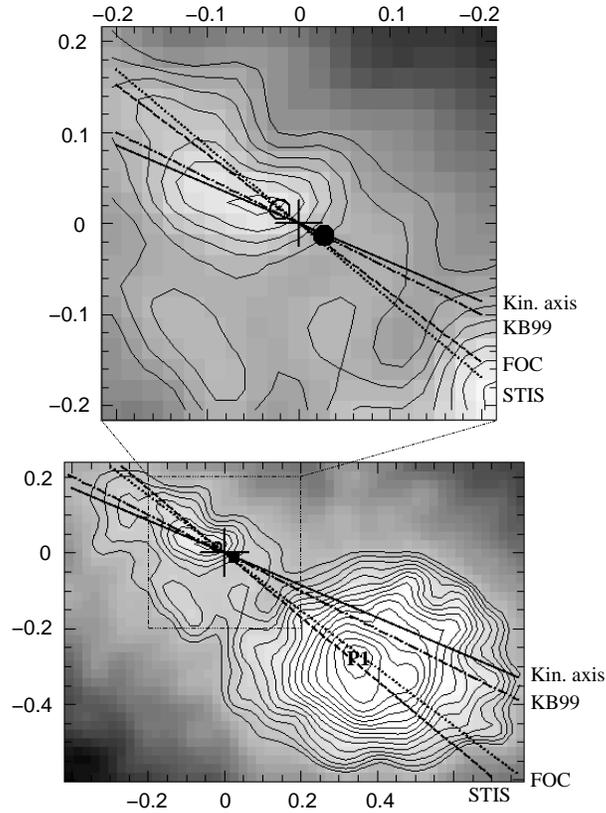,width=8cm}}
\caption{The central region of the deconvolved WFPC2 $I$ band
image, showing the locations of P1, P2, and the UV peak
(cross), as well as of the long-slits axes. The zero points
of the SIS and FOC long-slit data are indicated as a filled and empty
circle respectively. Note the significant difference between the
kinematical axis as measured with {\tt OASIS} and the STIS or FOC slits.}
\end{figure}

We have recently published new high resolution data of the nucleus
of M~31, obtained with the adaptive optics assisted integral
field spectrograph {\tt OASIS} (Bacon et al. 2001), as well
as archival STIS kinematics. 
Taking advantage of the two-dimensional spatial coverage
of the {\tt OASIS} data, and helped by multi-band dithered HST/WFPC2
photometry, we have reconsidered the centering of the
various published data sets. Using integral field spectroscopy,
we can accurately define a reference centre
{\em a posteriori}. Long-slit data of M~31 were alternatively
taken along the P1-P2 axis, and along the kinematical major-axis 
of the nucleus, these axes differing by more than 10 degrees.
Centres were also not necessarily well registered, and significant
shifts between the different data sets were measured.
These two point were important enough to confuse the interpretations, 
particularly during the modeling process (see Bacon et al. 2001 for details).

As emphasized in Fig.~1, it is thus not clear where a slit should be put.
The fact that P1 does not lie on the nucleus kinematical major-axis
must be taken into account when comparisons with models are performed.
Only high resolution two-dimensional spectroscopy coupled
with accurate photometry did allow a consistent picture to emerge.
\begin{figure}[h]
\centerline{\psfig{figure=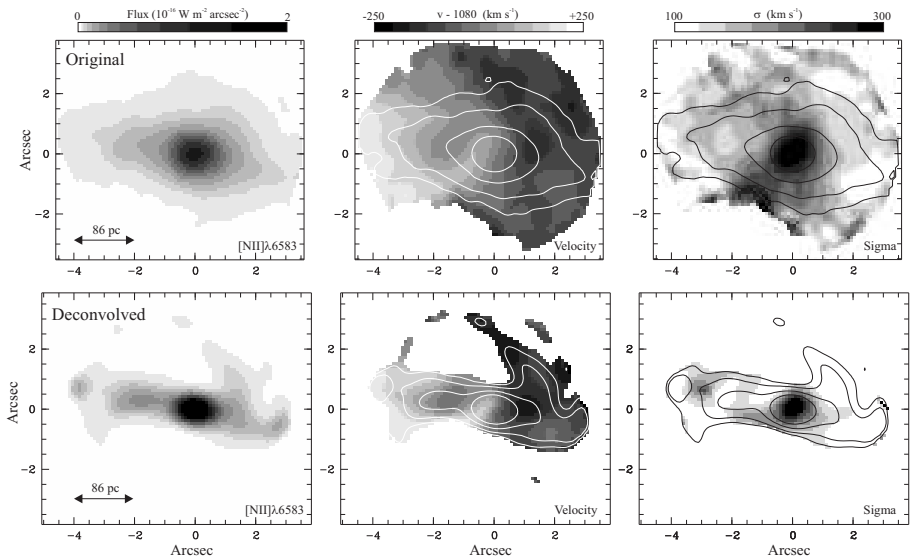,width=12cm}}
\caption{Original (top panels) and deconvolved (bottom panels) {\tt TIGER}
maps of the gas distribution (left, [NII]) velocity (middle) and dispersion (right).
See Emsellem \& Ferruit 2000 for details.}
\end{figure}

\subsection{The Sombrero: a double barred galaxy?}

The Sombrero galaxy (M~104) is another famous and well studied
nearby spiral. It is close to edge-on ($i = 84\deg$), which tends
to hide any departure from axisymmetry within its equatorial plane.
M~104 has been (and is still) assumed to be an atypical (because
of its huge spheroidal component) but axisymmetric early-type spiral.
Long-slit spectroscopy has thus often been used to probe its major or minor-axis
kinematics and line-strength gradients (e.g. Hes \& Peletier 1993).

However, the presence of an outer prominent ring of HI, CO and dust, 
led Emsellem (1995) to suggest the existence of a tumbling potential,
thus linking the ring to its Outer Lindblad Resonance. This was later
confirmed by the discovery of an additional Ultra Harmonic ring at a radius
of 43\arcsec (Emsellem et al. 1996). But the real
surprise came from the the {\tt TIGER} integral field 
spectroscopic data (Fig.~2). The two-dimensional coverage allowed us to 
spatially {\em deconvolve} the full {\tt TIGER} datacube
(reaching an equivalent resolution of $\sim 0\farcs5$),
something which would have been impossible using long-slit data.

The resulting emission line distribution shows the presence 
of a spiral-like structure, and the gas velocity map exhibits
strong non-circular motions within the few central arcseconds 
(Emsellem \& Ferruit 2000). We also detected the inner gaseous spiral 
in the narrow band WFPC2/HST ([NII]/H$\alpha$) imaging 
(see also Pogge et al. 2000), and a puzzling dust lane system
using HST colour maps (Fig.~3). At a minimum inclination of 82\deg, 
the deprojection stretching factor along the minor-axis is
about 6. If we then assume that the detected gas 
lies within the equatorial plane of the galaxy, its distribution
must therefore be significantly elongated. Hence we tentatively suggested the presence
of an inner bar, the Sombrero galaxy being then a {\em double barred spiral}.
\begin{figure}
\centerline{\psfig{figure=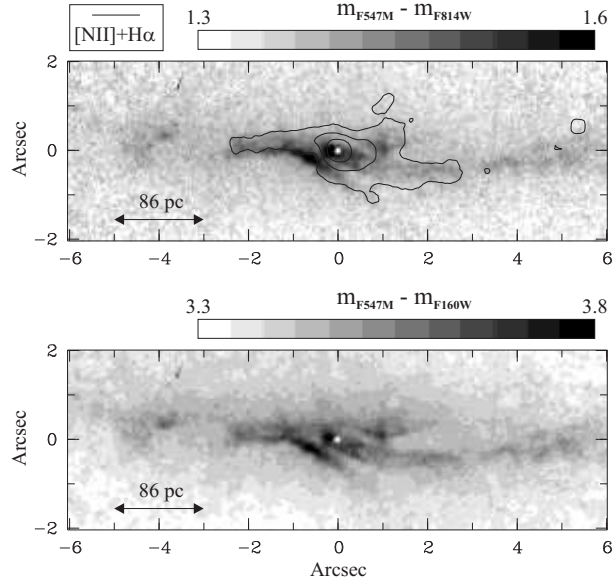,width=8cm}}
\caption{Top: $V-I$ HST colour map, with some isophotes of the 
HST [NII]+H$\alpha$ image superimposed. Bottom: $V-H$  HST colour map.}
\end{figure}

\section{Time evolution}

As mentioned in the introduction, density waves in the inner regions
of galaxies imply a strongly time dependent morphology. This can for example
be shown by numerical simulations with which we can 
take snapshots of observable quantities at different times.
\begin{figure}
\centerline{\psfig{figure=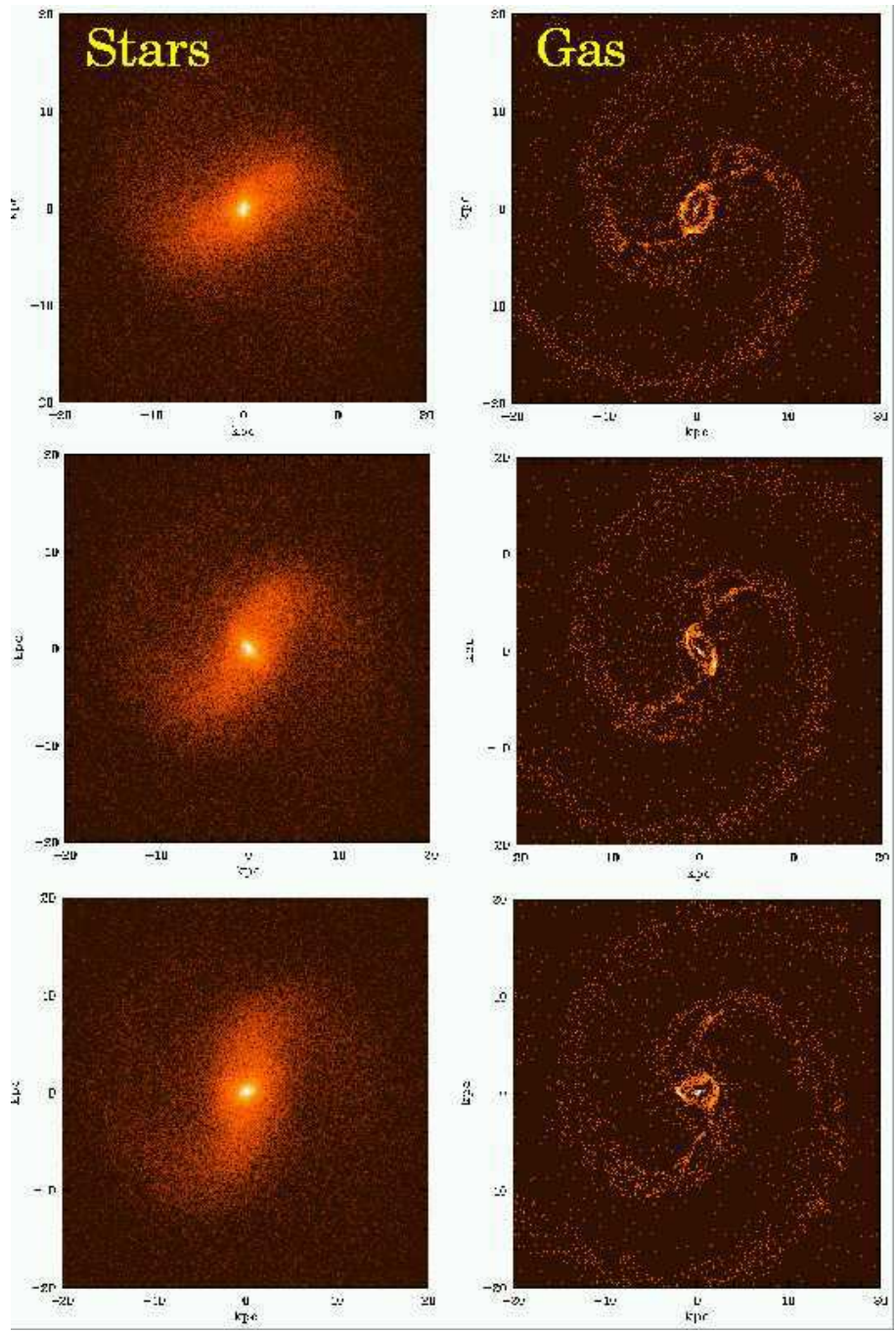,width=12cm}}
\caption{Snapshots of a N body $+$ SPH simulation for a double barred
galaxy (courtesy of Daniel Friedli). Face-on view of the stellar (left)
and gas (right) components. Time goes from top to bottom.}
\end{figure}

\subsection{Following a double bar system}

Double bars are ideal cases to illustrate the issue of time dependency,
as the two bars are generally thought not to share the same pattern speed.
This implies a strongly evolving dynamics and morphology particularly
at the interface region, as nicely emphasized by Maciejewski et al. (2002).
A N-body$+$SPH simulation of a double barred galaxy (kindly provided 
by Daniel Friedli) is presented in Fig.~4. In this model, the inner
bar rotates faster than its large-scale parent: this can easily be
seen by tracing their position angle difference. The gas distribution
at and within the Inner Lindblad Resonance ring
exhibits a strongly {\em evolving} shape. 

In real galaxies, these structures are projected on the sky.
Strong departures from circular motions will have to be interpreted
in the frame of a deprojected morphology. Only a full two-dimensional
view can reconcile the observations with the models, and help
constraining the true morphology and dynamics of the target.

\subsection{More ingredients}

The model of a double bar galaxy presented in the previous Section (Fig.4),
did not include processes such as star formation or nuclear activity.
When these ingredients are added, things get even more confused.
Cases like NGC~1808 (Emsellem et al. 2001) are there to remind us that there are 
important couplings between various intervening processes which
may lead to complex cycles (see e.g. Combes 2001). 

I present here an example of such a complex case, namely
the double barred galaxy NGC~470. It is a star forming inclined spiral,
the photometry of which looks rather regular in the near-infrared 
(Friedli et al. 1995). We recently obtained {\tt OASIS} integral field observations
of the central region of NGC~470 (Fig.~5). The gaseous and stellar distribution
in the optical are strongly asymmetric. This is partially due to the presence
of dust on the NW part, but there seems to be a true residual 
lopsidedness in the central few arcseconds. Without these maps, it would
have been difficult to determine where the kinematical centre lies.
\begin{figure}
\centerline{\psfig{figure=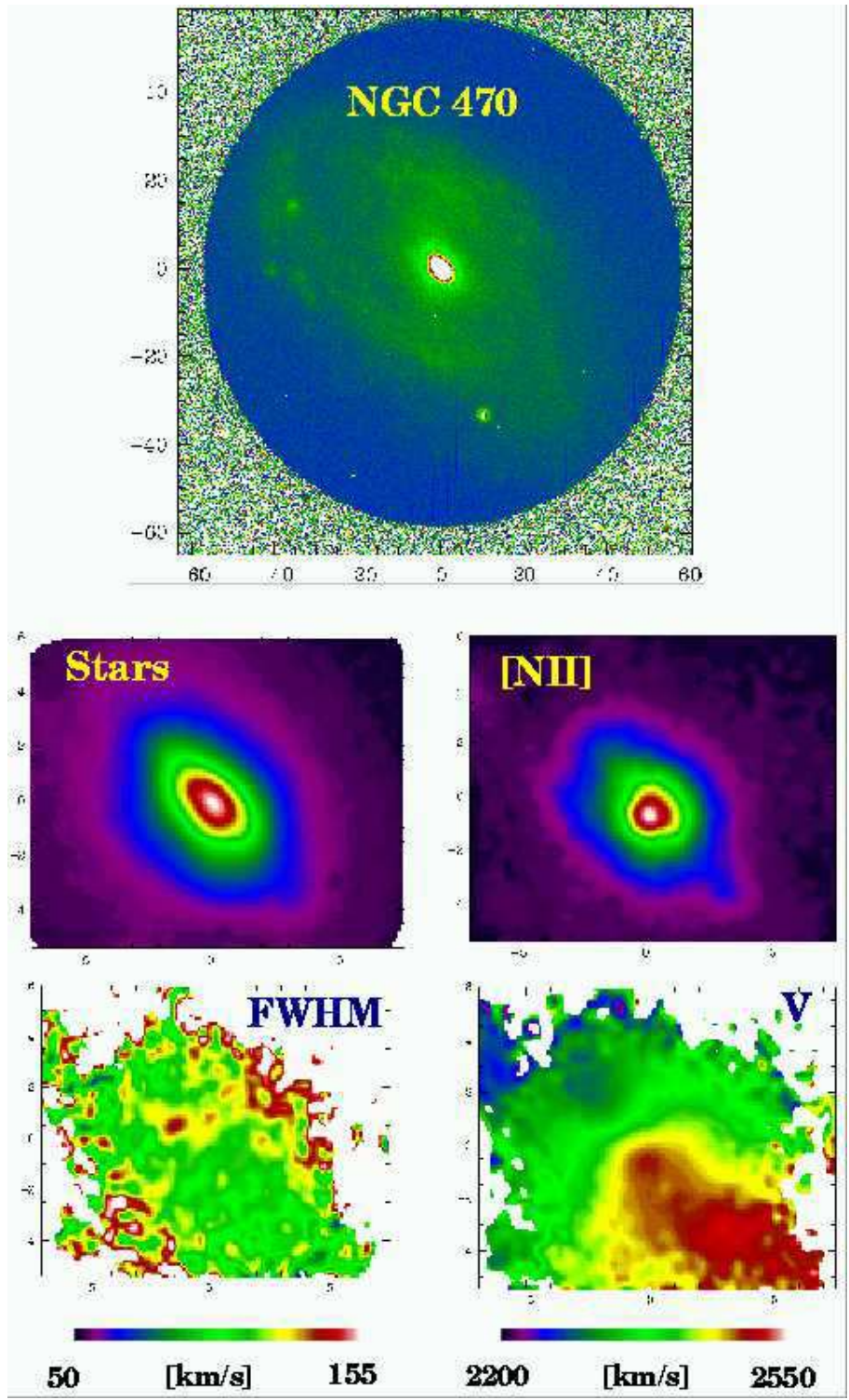,width=12cm}}
\caption{{\tt OASIS} observations of NGC~470. Top panel: direct image.
Middle panels: stellar (left) and gas (right) distributions.
Bottom panels: gas velocity (right) and dispersion maps (left). Note
the offcentring of the kinematical centre and dispersion peak.}
\end{figure}

\section{Conclusions}

Density waves seem to be ubiquitous in the central regions of galaxies, 
although we still do not have sufficient observational constraints
to assess their existence. If they do exist, density waves 
certainly play an important role in redistributing
the dissipative component and reshaping the inner parts of disk galaxies.
Two-dimensional spectrographs are now becoming more common, and every
large telescope will have its own. With their enhanced spatial resolution
and the parallel increase in collecting aperture, these will be ideal instruments 
to probe the existence and role of inner density waves.

\end{document}